# Thermal Preconditioning of Membrane Stress to Control the Shapes of Ultrathin Crystals

Hao Wan,[1] Geunwoong Jeon,[2] Gregory M. Grason,[1] Maria M. Santore[1,*]

[1]Department of Polymer Science and Engineering
University of Massachusetts
120 Governors Drive
Amherst, MA 01003

[2]Department of Physics
University of Massachusetts
710 N Pleasant St
Amherst, MA 01003

*corresponding author: santore@mail.pse.umass.edu

## Abstract

We employ the phospholipid bilayer membranes of giant unilamellar vesicles as a free-standing environment for the growth of membrane-integrated ultrathin phospholipid crystals possessing a variety of shapes with 6-fold symmetry. Crystal growth within vesicle membranes, where more elaborate shapes grow on larger vesicles is dominated by the bending energy of the membrane itself, creating a means to manipulate crystal morphology. Here we demonstrate how cooling rate preconditions the membrane tension before nucleation, in turn regulating nucleation and growth, and directing the morphology of crystals by the time they are large enough to be visualized. The crystals retain their shapes during further growth through the two phase region. Experiments demonstrate this behavior for single crystals growing within the membrane of each vesicle, ultimately comprising up to 13% of the vesicle area and length scales of up to 50 microns. A model for stress evolution, employing only physical property data, reveals how the competition between thermal membrane contraction and water diffusion from tensed vesicles produces a size- and time-dependence of the membrane tension as a result of cooling history. The tension, critical in the contribution of bending energy in the fluid membrane regions, in turn selects for crystal shape for vesicles of a given size. The model reveals unanticipated behaviors including a low steady state tension on small vesicles that allows compact domains to develop, rapid tension development on large vesicles producing flower-shaped domains, and a stress relaxation through water diffusion across the membrane with a time constant scaling as the square of the vesicle radius, consistent with measurable tensions only in the largest vesicles.

**Introduction**

The development of thermal stress is critical in the production and performance of ultrathin films and engineering materials, yet stress evolution can be just as important in less expected situations. These include nanometrically thin supported phospholipid bilayers[3-7] or vesicles of the same.[1, 8-13] Giant unilamellar vesicles (GUVs) of phospholipid bilayers are most often studied to enable insights into biological systems; however, GUVs and liposomes also contribute materials relevance in their potential to produce ultrathin crystals,[1, 14-16] delivery packages,[17, 18] and functional skins.[19-22] Here we address how the elasticity and curvature of vesicle membranes, compounded with stresses from thermal processing, produce a complex and temporally variable environment for 2D crystal growth, dramatically impacting crystal morphology. We demonstrate how process history can be exploited to control membrane tension, so that membrane bending selects for different crystal morphologies on vesicles of different size.

Fluid ($L_\alpha$) phospholipid bilayers are 2D elastic fluids, able to flow (shear) in plane while paying an elastic cost to bend and stretch.[23, 24] With lamellar regions that are composed predominantly of hydrocarbons, phospholipid bilayers possess coefficients of thermal expansion on the order of 1-5 x $10^{-3}$/°C.[23] Supported fluid phospholipid bilayers experience stress as a result of substrate interactions and differences in thermal expansion coefficient with substrates of different materials.[7, 25] Less obvious, thermal contractions of phospholipid bilayer membranes in vesicles can be limited by the interior water. This constraint gives rise to membrane stress upon cooling, central to this work.

Figure 1 shows how the thermal contractions and effective area reduction of a phospholipid vesicle occur upon cooling from temperature near 45°C. (The composition, 100 % DOPC, produces a membrane that is in the $L_\alpha$ fluid phase for the full range in this demonstration.) A flaccid vesicle has been selected to clearly reveal membrane contraction. This qualitative experiment involves recording the fluctuating vesicle shape of the un-tensed vesicle via fluorescence video microscopy and then placing a drop of ice-water on the coverslip. The vesicle immediately "shrinks" and appears round in cross section, reflecting the overall spherical shape that the vesicle has assumed, a result of thermal membrane contraction. The membrane tension is initially near zero when the membrane undulates but, from the instant when the GUV first becomes spherical, any further cooling causes an increase in tension due to the further reduction what would be the in the zero tensioned membrane area and the larger area needed to encapsulate the water inside the vesicle. If the tension becomes too high, the membrane can rupture or lyse. Such rupture does not always destroy the vesicle. In many cases, membrane pores reseal after some loss of contents, reducing tension.[26, 27] Tension will continue to rise with further cooling

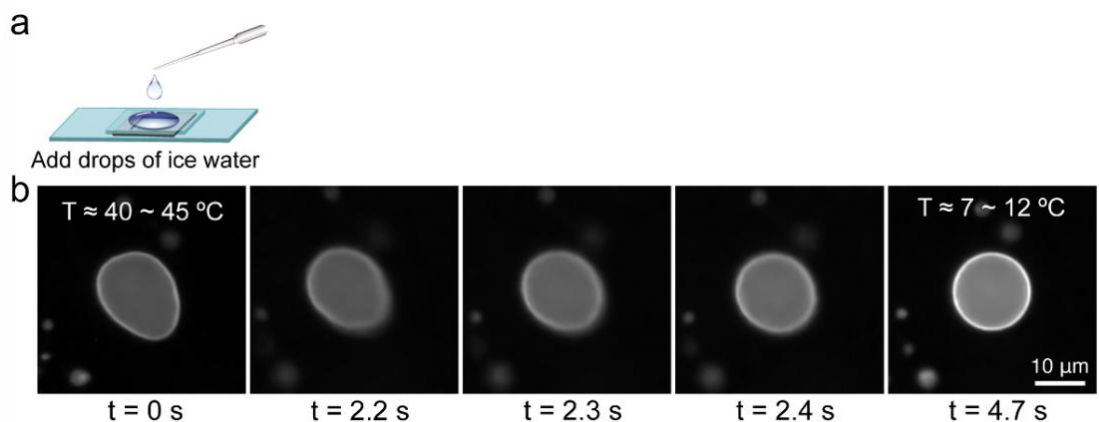

**Figure 1**. (A) Schematic of qualitative experiment: Ice water is placed on top of a coverslip which is part of a sandwich containing a suspension of fluorescently labeled DOPC vesicles initially at elevated temperature, 40-45°C. (B) Image series of a vesicle that was initially flaccid at elevated temperature. Upon rapid cooling (temperature not recorded), the vesicle membrane contracts and within seconds the vesicle becomes spherical.

and the burst-reseal process can be cyclical in vesicle volume, membrane areal strain, and tension. A similar burst-reseal effect has been established employing osmotic stress.[28, 29]

The thermal stress development implied in Figure 1, for vesicles whose membranes are entirely $L_\alpha$ fluids, would be expected to substantially affect the growth of any crystalline solid membrane phase developing during cooling. In this work we consider the crystallization of 1,2-dipalmitoyl-sn-glycero-3-phosphocholine (DPPC, $T_m$= 41°C), by nucleation and growth within mixed membranes of DPPC and 1,2-dioleoyl-sn-glycero-3-phosphocholine (DOPC).[30, 31] The membrane-integrated DPPC crystals formed on cooling are nearly pure[8, 9, 32] and possess similar order and molecular packing on both the inner and outer bilayer leaflets,[33-36] imparting a preferred flatness to the essentially 2D crystals. [1][1, 37-40] The DPPC crystals are distinguished from the $L_\alpha$ fluid phase primarily by their shear rigidity[23, 24] though, like the $L_\alpha$ fluid membrane, the solid crystals also pay an elastic bending cost.[41]

We previously reported a correlation between vesicle size and the shape of single solid crystals grown within vesicle membranes.[1] The crystal shapes spanned from hexagons to flowers with increasingly long protrusions or petals in increasingly larger vesicles. In contrast, crystallization on a rigid spherical template in the absence of a membrane exhibits the opposite effect: greater protrusions, stripes, and petals on templates of sharper curvature (smaller spherical templates), a result of stress sustained within crystals which resist taking on Gaussian curvature.[42-45] The difference between the two cases suggests that in vesicles, the fluid membrane bending energy can dominate that of the solid in directing the growing crystal morphology. Our Surface Evolver computations confirmed the importance of the fluid bending energy, showing the preference for flower-shaped crystals over compact hexagons in inflated high -tensioned vesicles.[1] We also experimentally confirmed higher tensions in the flower-producing vesicle, which happened to be the larger ones in a given sample. Thus, with strong influence of the fluid membrane on the crystallization of a solid phase, different from crystallization on a rigid template, manipulation of the membrane could provide a means of tuning crystallization.

In this work we explore the thermal "preconditioning" of mixed phospholipid vesicles in the one-phase fluid region of the phase diagram as a means to manipulate membrane stress prior to initiation of crystallization. We expect that variations in preconditioning will shift the relationship between vesicle size and crystal morphology. Using membranes containing mixtures of DPPC and DOPC as a model, we target conditions that produce a single growing crystal within each vesicle membrane in order to most clearly elucidate the impact of cooling rate and membrane tension on 2D crystal morphology. We combine our experimental findings with a model predicting the membrane tension history upon preconditioning in the one phase region, revealing how vesicle size influences tension and pre-determines early crystal shape. Thus, in this work we demonstrate how preconditioning membrane tension, through thermal processing, enables control on the crystal shapes that are produced in vesicles of different sizes and how this can work for hundreds of vesicles / crystals at a time.

**Results and Discussion**

*Features affecting morphology*

Vesicles were produced by electroformation at an elevated temperature in the one phase region of the phase diagram to impart compositional vesicle-vesicle uniformity. After harvesting and storage typically for up to two days but never more than three, vesicle suspensions were reheated to the one phase region of the phase diagram, annealed for a timescale of minutes to ensure uniform mixing of lipids within each vesicle's membrane, and then controllably cooled into the two phase region that starts near 31°C in Figure 2,[1] to initiate the formation of membrane-integrated DPPC crystals by nucleation and growth. Ultimately vesicle suspensions reached room temperature. The composition of 30 wt% DPPC / 70 wt% DOPC, on the DPPC-dilute side of the phase diagram, favors the nucleation of a single solid crystal within each vesicle,[1, 22, 31] especially at slower cooling rates. With larger vesicles and faster cooling rates, multiple domains were both expected and observed on some vesicles but, for consistency, not included in the study.

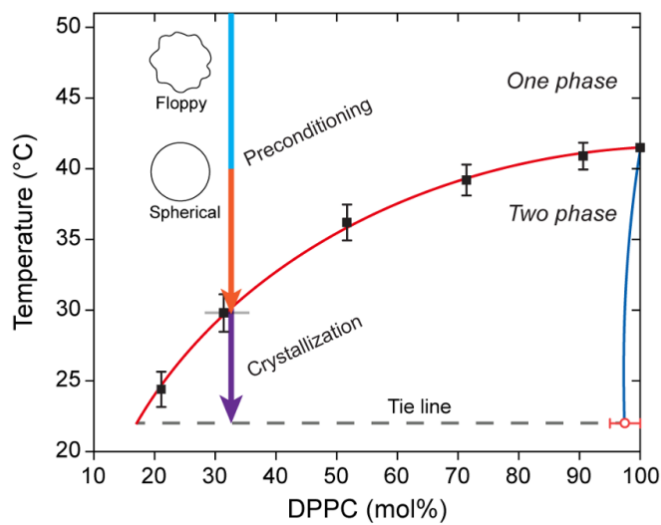

**Figure 2**. Phase diagram (data from reference[1]) for DOPC/DPPC mixed vesicles, as measured by fluorescence. Annotation indicates the thermal path to produce the crystals, including an approximation of where floppy vesicles, at the highest temperatures, become spherical on cooling, still in the one phase region.

The images in Figure 3 show, for two different vesicles cooled at 1.2°C/min, the progressive growth of flower- and hexagon-shaped crystals, which are the dark regions, excluding a tracer dye which remains in the fluid phase. The six-fold symmetry is typical of the DPPC crystals. It is particularly striking that the solid crystalline domains, at the time when they can first be optically resolved, have already taken their ultimate shape, in these cases flowers or hexagons. With further cooling and crystal growth, the established domain shapes simply grow larger.

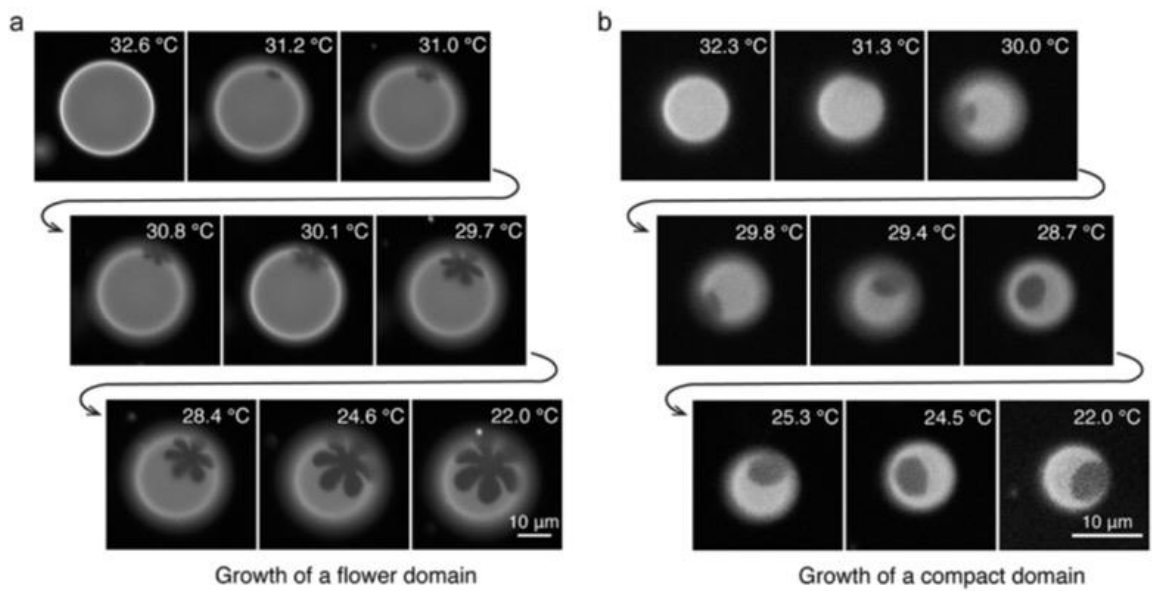

**Figure 3**. Growth of crystals during cooling through the two phase region for a) a flower shaped crystal and b) a hexagonal crystal. Vesicles were cooled at 1.2°C/min.

The progressive growth of the initially-formed shapes in Figure 3 is typical of all the vesicles in this study, independent of cooling rate, and suggestive of molecular addition at the crystal edges.

Similar early commitment to crystal shapes for 0.30°C/min have been previously documented in micrographs.[1] In particular, the growth of flower shapes demonstrates that they are not produced by aggregation of smaller domains or dendrite-type propagating instabilities. Also important for the vesicles in Figure 3 near room temperature and, more generally for all the vesicles in this study when observed at room temperature, the solid area fraction on each vesicle was effectively constant, near 13± 2.5 %, a result of the fixed overall composition of all the vesicle membranes. This result is in agreement with a reading of the molar proportions of fluid and solid phases on the phase diagram[1, 30] plus separate data for the molecular areas within fluid and solid phases.[24, 46] Thus the difference between the different crystals is simply their shape. Findings at 0.30°C/min, together with Figure 3, suggest similar crystallization and growth mechanisms at different cooling rates.

Beyond the early commitment to a given crystal shape upon entry into the two phase region, a second distinguishing feature of all the membrane crystals in this work is the correlation between vesicle size and the shape of the crystal. This is evident in Figure 4 for each of three cooling rates. Compact crystal shapes such as hexagons are seen on smaller vesicles for a given cooling rate, while starfish and flower shapes are seen on vesicles of increasing diameter. While the ratio of the overall (outer) crystal radius to its inradius is a continuum, certain ranges of shapes, described by ranges of this ratio, alpha in Figure 4b, are visually distinguishable. These include hexagons, starfish, simple flowers and flowers with extended or even serrated flowers. This fact enables the creation of histograms relating crystal shape to vesicle size.

In the histograms of Figure 4, vesicles were categorized based on the shapes of their crystals and then the vesicle diameter was measured to produce a data point. Each histogram in Figure 4, describing the crystal shape-vesicle size distribution, was measured for three independent runs at a select cooling rate, comprising 175-375 vesicles in each histogram. Comparison of the three rates, 1.2 °C/min (70 °C/hr), 0.30 °C/min (20 °C/hr), and 0.013 °C/min (0.8 °C/hr), shows that the relationship between vesicle size and crystal shape is shifted by cooling rate. Slow cooling rates favor more compact crystals for the size range of vesicles we are able to produce. With very few vesicles exceeding 50 μm diameters, the slow cooling reduces the range of alpha to less than 2.4. Thus slow cooling produces a smaller range of crystal shapes on these vesicles. For very fast cooling, 1.2 °C/min, a greater fraction of the vesicles display flower-shaped crystals of varying petal lengths.

All the data in Figure 4 can be summarized in the state space of Figure 5A, which maps cooling rate and vesicle size to the shape of the crystal that was observed room temperature. Figures 4 and 5 both reveal that increased cooling rate shifts the preference for flower-shaped crystals to small vesicles, while slow cooling increases the range of vesicle sizes on which compact or hexagonal crystals can be grown. Also, faster cooling rates seemed to reduce our ability to locate large vesicles possibly due to breakage. These findings suggest that an increased cooling rate may generally elevate vesicle tension, shifting the preference for flower-shaped crystals to a larger swath of processing space, explored via modeling, in the next section.

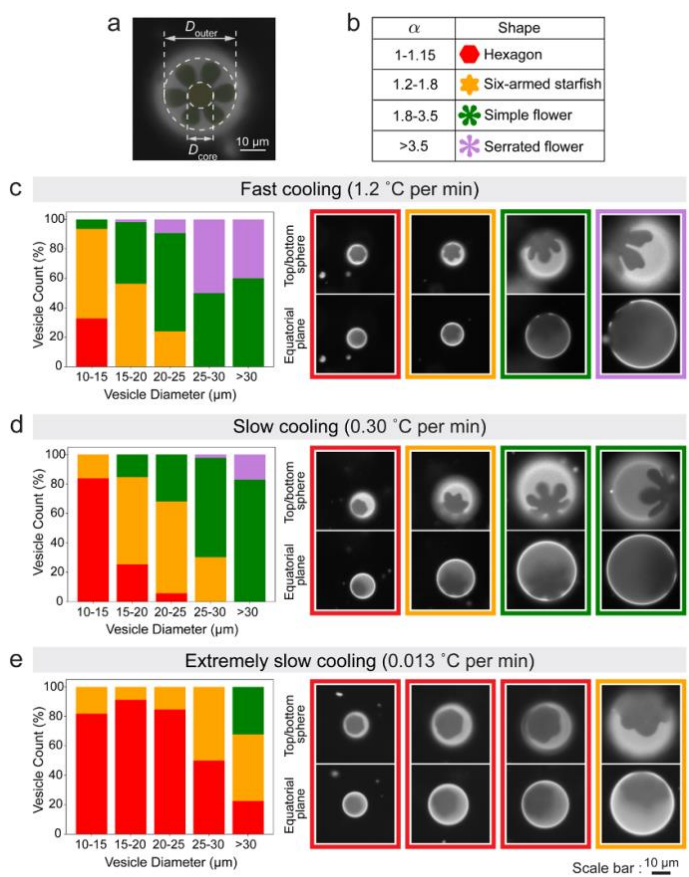

**Figure 4**. a) Illustration showing measurement of outer and inner crystal diameters, the ratio of which gives alpha. b) Range of alpha values for different pattern types. c-e) Histograms showing relationship between pattern type and vesicle size for different cooling rates and, for each cooling rate, examples of the different patterns on vesicles of different sizes.

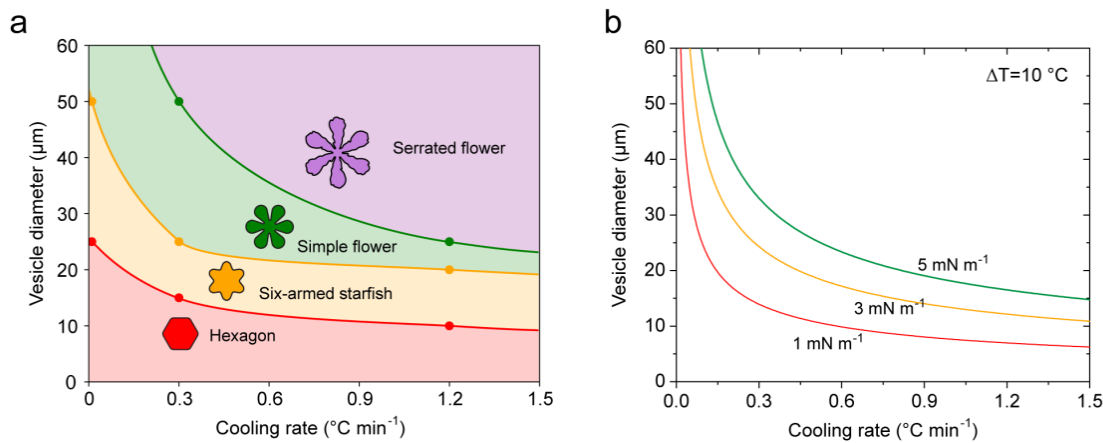

**Figure 5**. a) State space, based on the mid-points of the histogram distributions in Figure 4, showing the patterns formed for vesicles of different sizes and cooling rates. The points represent the vesicle sizes corresponding 50% in the distribution separating each shape category, based on Figure 4. The full distributions for the vesicle size-crystal pattern relationship, along with error bars, can be found in the Supporting Information. b) loci of membrane tension predicted by the engineering model for vesicles of different sizes and subject to cooling at different rates for a ten degree temperature drop.

*Calculating Evolution of Tension During Cooling.*

Given that the crystal shapes seen at room temperature grew from similar shapes seen shortly after nucleation, and knowing the that the costs of membrane bending is amplified when vesicles are tensed,[1, 22] we considered the evolution of tension during cooling. Membrane tension is controlled by two completing processes: thermal contraction which tends to increase membrane tension upon cooling vesicles that are filled to capacity with an aqueous solution, and the pressure difference across the membrane that drives water from tensed vesicles and tends to reduce tension. Here we develop a model of these two competing effects to estimate the vesicle-size dependent evolution of membrane tension when vesicle suspensions are cooled to nucleate and grow crystals. For membrane compositions of 30% DPPC 70% DOPC entry into the two phase region occurs near 31°C, but vesicles are typically annealed at higher temperatures explained in the methods section.

As established in Figure 1, when a vesicle is cooled, if it is floppy, its fluid membrane contracts far more than the aqueous contents, and it ultimately reaches a spherical shape. During cooling, as long as the vesicle is floppy, its tension approaches zero. Then the area, $A$, follows the equilibrium value, $A_{eq}$, prescribed by the coefficient of thermal expansion, $\kappa$:[23]

$$\kappa = \left[\frac{1}{A_{eq}}\frac{\partial(A_{eq})}{\partial T}\right]_{\tau=0} \tag{1A}$$

If a cooling *rate* is applied, then the zero tension state follows:

$$\left(\frac{dT}{dt}\right)\kappa\, dt = \frac{dA_{eq}}{A_{eq}} \tag{1B}$$

While the tension vanishes and the vesicle is floppy, $A = A_{eq}$. Once the vesicle is spherical, however, further cooling reduces the equilibrium area according to equation 1. The actual area $A$ is, however, constrained by the vesicle contents, and the tension, $\tau$, is finite:

$$\tau = K_a\left(\frac{A-A_{eq}}{A_{eq}}\right) = K_a(A' - 1) \tag{2}$$

Here, $K_a$, is the area expansion modulus and $A'$ is the areal deformation, $A/A_{eq}$.

When the vesicle membrane is tensed, the pressure inside the vesicle exceeds the exterior pressure as described by LaPlace:

$$\Delta P = \frac{2\tau}{R} \tag{3}$$

where R is the vesicle radius. Since phospholipid membranes are somewhat water permeable, this pressure difference from the inside to the outside of the vesicle drives water out of the vesicle, and produces a volume reduction:

$$\frac{dV}{dt} = -A \frac{\wp}{\rho} \Delta P \qquad (4)$$

Equation 4 defines the membrane permeability, $\wp$. Also, $A$ is the vesicle membrane area, and $\rho$ is the density of water.

For vesicles that are spherical, $V$ and $A$ are related such that $dV = \frac{1}{4\sqrt{\pi}} A^{1/2} dA$. With this, equations 1B through 4 can be combined to yield:

$$dA' = -\left[k''(A'-1) + A'\left(\frac{dT}{dt}\right)\kappa\right] dt \qquad (5)$$

This form explicitly shows the impact of transport (the first term in square brackets) and thermal contraction (the second term in square brackets). The constant

$$k'' = 16\pi \frac{\wp}{\rho A_{eq}} K_a, \qquad (6)$$

combines the membrane's water permeability and area expansion modulus. It also contains, grouped with the permeability, the equilibrium area of the membrane (available for water transport), which changes with cooling by a few percent. It should be noted, however, that changes in $A_{eq}$ at this position of the equation have a minimal influence on its solution, no more so than minor (a few percent) changes in the permeability or area expansion coefficient, which are treated as constant. The impact of cooling on these physical property data, or $A_{eq}$ of that matter, would constitute a further refinement of the model and so the quantities are grouped, in the current treatment into constant $k''$. The analytical solution to equation 5 is

$$A'(t) = \frac{k''}{k'' + \kappa\frac{dT}{dt}} + \left(\frac{-k''}{k'' + \kappa\frac{dT}{dt}} + A'_{t=0}\right) exp\left[-\left(k'' + \left(\frac{dT}{dt}\right)\kappa\right)t\right] \qquad (7)$$

The evolution of membrane tension can be further calculated from *A'(t)* employing equation 2. Equation 7 can be solved with no adjustable parameters, employing estimates for the physical property data from the literature.

*Tension Evolution at Experimental Conditions*

Figure 6A illustrates how, for spherical vesicles of different sizes initially at zero tension, tension increases as a result of cooling, here with a cooling rate of 0.30 °C/min, corresponding to the intermediate cooling rate in Figure 4. Other physical properties from the literature, corresponding to our DOPC-DPPC fluid-phase vesicles were employed: $\kappa = 0.005$ K$^{-1}$,[23] $K_a = 237$ mN/m;[47] $\wp = 42 \frac{\mu m}{s}$;[47] and $\rho = 1$ g/cm$^3$.[2] In Figure 6, the calculations match a typical cooling run: cooling from an elevated temperature for some period on the order of an hour and then holding at a final temperature, for instance room temperature. The value of temperature is not, itself, a fundamental part of the calculations; instead, the model shows the evolution of tension starting with $\tau = 0$ from the point where the vesicle becomes spherical. This might occur near 40°C, the example indicated by the temperature scale at the top of Figure 6A. The cooling portion of the calculation is followed by a calculation for constant temperature, matching experiments. This is achieved in equation 7 by setting *dT/dt = 0* after the period of cooling.

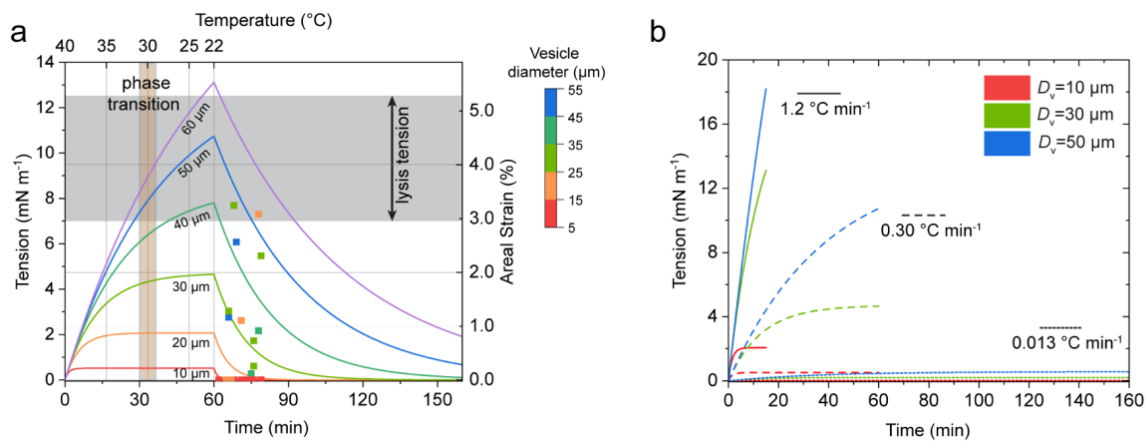

**Figure 6**. a) Predictions for the variation of tension in time, starting with spherical vesicles at zero tension, for vesicles of different sizes, subject to cooling at 0.30°C/min for 1 hour and then held at constant temperature. $\kappa$= 0.005 K$^{-1}$, K$_a$ = 237 mN/m, $\wp = 42$ µm/s. The temperature scale on the upper x-axis shows how temperature would evolve if vesicles first became spherical near 40°C, so that when subjected to further cooling their tensions would increase. The region where the system crosses into the two phase region is indicated. Lysis conditions (9.9 ± 2.6 mN/m)[2] are also indicated. b) Predictions from the engineering model for the impact of different cooling rates on the tensions in vesicles of different sizes, indicated.

Figure 6A reveals striking and unanticipated general features of the tension evolution. First, the vesicle size exerts a dramatic influence on the tension and the very character of tension evolution. Beyond the expected low tensions in small vesicles, (expected for water diffusion across a larger area of membrane per unit interior volume in small vesicles compared with larger ones), small vesicles achieve a low steady state tension within minutes of cooling while large vesicles experience a protracted large tension increase over the entire hour-long cooling period. Small vesicles, here about 20 µm or smaller, never achieve an appreciable tension and instead reach a low plateau level. Conversely, the stresses and strains on vesicles exceeding 40 µm quickly become substantial and exceed lysis conditions.[2] This qualitative difference in behavior can be understood because of the effectively greater water loss across the membrane per unit vesicle volume in small versus large vesicles. Small vesicles cannot hold their water contents effectively and reach a plateau, much like trying to pump up a leaky tire, which retains only partial inflation. This difference is a result of the volume to surface area ratio and not an issue with membrane integrity in smaller vesicles.

When cooling ceases and the temperature is held fixed at room temperature, the membrane contracts towards the fixed value of $A_{eq}$, enabling relaxation of membrane tension. On small vesicles, the tension relaxes to zero within minutes while tension relaxation on large vesicles is incomplete on the timescale of interest, leaving measurable residual stresses persisting for more than 30 minutes. In the absence of cooling, the tension relaxation is exponential, with a time constant of $k''^{-1}$, giving a radius-squared timescale for approach to zero tension.

*Impact on Crystallization*

The behavior in Figure 6 informs our understanding of the states of vesicles while still in the one-phase region, important in explaining the predisposition of crystals to take on compact or hexagonal shapes in Figure 3. While we only approximate when vesicles first become spherical and develop finite tension, the vesicles do appear round for ~30 minutes during the approach to the two phase region near the cooling rate of 0.30°C/min. Therefore we might guess that many of the vesicles are spherical at the time the temperature is ~40°C, which allows for tension to develop during ~10 Celsius degrees of cooling prior to reaching the two phase region. Were vesicles to start their increase in tension at 35° rather than 40°C, then one would reach the two phase region sooner on the time axis of Figure 6. This would have negligible impact on the tension of the small vesicles, and the larger vesicles would simply not be preconditioned for as long, achieving somewhat lower tensions at the time of nucleation, when 31°C is reached.

Once inside the two phase region, there is between 8-10°C of additional cooling before the end of the experiment at room temperature. The tension evolution during this period is expected to be similar to that during the preconditioning of the fluid vesicle except, in the two phase region, a small section of membrane, growing progressively from 0 to 13 % of the total area, possesses the physical properties (molecular area, Ka, $\wp$, and κ) of the solid rather than the fluid membrane. This can change the ultimate tension evolution only modestly, for instance if the crystal were 15% smaller than the fluid membrane in molecular area and impermeable to water. The correction to the tension will become most important at room temperature when the crystal reaches 13% of the membrane area, but in the initial stages of growth, the small area of the crystal limits the impact of the solid properties on the tension into the two phase region. The impact of overall membrane properties is explored in the supporting information and is most pronounced for the largest vesicles. Any corrections to the tension evolution in the two phase region would have no impact on the tension preconditioning that occurs in the one phase region and the tension environment experienced by small crystals.

*Exploiting Membrane Preconditioning to Manipulate Crystal shape.*

Figure 6 is remarkable in its prediction of high and low tension regimes, which mostly agree with the tensions measured by micropipettes. These previously reported values,[1] incorporated here alongside calculations, were quantified using micropipettes after vesicles were cooled from the one phase region to room temperature 0.30°C/min. After reaching room temperature, the vesicle suspension was transferred from the closed thermal chamber to an open chamber accommodating micropipettes, the transfer process taking 5-10 minutes. Then only 1-4 vesicles could be immediately measured (in the next 10 minutes, for a total time of 20 minutes at room temperature). and with the clock ticking, further vesicles were aspirated. The experiment was repeated in order to gather sufficient data. In all cases except two vesicle, for 8 runs and 29 vesicles, those with diameters less than 25 μm never registered a measurable tension, in excellent agreement with the predicted tension evolution in Figure 6A. The two 25 μm vesicles (orange points) measuring a finite tension had pronounced flower-shaped crystals, suggesting they were in the upper tension range for vesicles of their size. Finite and elevated tensions were measured on larger vesicles which were fewer in number, possibly a result of breakage in those runs.

Figure 6B shows the impact of cooling rate on the evolving tensions experienced by vesicles of different sizes: 10, 30, and 50 μm in diameter. It is interesting that the intermediate rate spreads the tension curves substantially during a useful laboratory time frame. The fast cooling causes vesicles larger than ~30 μm in diameter to rapidly develop tension as room temperature is approached, while only the smallest vesicles experience a plateau. The slow cooling rate of 0.013 °C/min produces plateaus in all three vesicle sizes and the approach to room temperature takes more than a day. These general trends, with rising tensions associated with flower shaped crystals and tension plateaus below 1mN/m for compact or hexagonal crystals are in excellent qualitative agreement with experiments.

In order to gain insight into the experimental state space of Figure 5A, the model was solved to identify, for a given cooling rate, the vesicle size that corresponded to a target tension. Then the loci of the predicted lines of constant tension are plotted in Figure 5B. The lines of constant tension match roughly with the divisions in the experimental state space between the different (arbitrary) classes of crystal shapes in Figure 5A. The only significant discrepancy occurs at the slowest cooling rates where the largest vesicle exhibited crystals with irregular (lobed) hexagons, which we identified, due to their concave character, in the starfish category rather than regular hexagons. Thus membrane tension, even before nucleation of crystals, directs the morphology of crystals once they are large enough to be visualized in the microscope.

*Why Tension Selects for Crystal Shape*

The underlying reason why crystal shapes with increasingly long protrusions (starfish arms, flower petals) are preferred on large vesicles that are more nearly inflated due to effectively slower water transport, lies in the ability of the protrusions to bend to accommodate larger relative vesicle volumes. Accommodation of more inflation by having petals that encircle the vesicle reduces the membrane tension which would otherwise have grown even larger, for instance if the crystal were flat. This point is made in Figure 7, which shows Surface Evolver calculations, detailed in our prior work[1] and summarized briefly in the current methods. Figure 7 compares three vesicles with differently-shaped membrane crystals. While the edges of the round domain buckle with locally cylindrical curvature, leaving a flat hexagonal center that contains much of the crystal area, the starfish arms and flower petals bend cylindrically over the rounded inflated vesicle. Though increases in internal volume will generally increase membrane tension, the flower-shaped crystal accommodates a greater relative internal volume compared with a flat disc-shaped crystal of equal area. This is represented, in Figure 7, by the relative volume, $\bar{v}$, the ratio of the actual vesicle volume to a sphere of equal area. Non-zero Gaussian curvature shapes like spherical caps, which would need to be assumed by a compact crystal on a spherical vesicle with trapped interior water, incur energetically prohibitive shear and stretching deformations.[48, 49] By contrast, when long protrusions such as flower petals bend cylindrically (and economically from the energy standpoint) around a vesicle, the non-zero Gaussian curvature that accumulates in the fluid portion of the membrane is much less energetically expensive due to a fluid membrane's ability for in-plane shear flow. Through this mechanism of protrusions on the crystal, a greater volume of water can be accommodated inside the vesicle before there is an increase in membrane tension. This, of course, comes at the cost of additional line energy between the fluid-crystal boundary, but as shown previously for sufficiently high inflation,[1] the lower elastic (bending) energy possible in non-convex solid domain shapes is more than sufficient to favor flower formation.

The curvature distribution between fluid and solid domains on equilibrated vesicles, including lack of Gaussian curvature in the solid, is evident in the Surface Evolver modeling results of

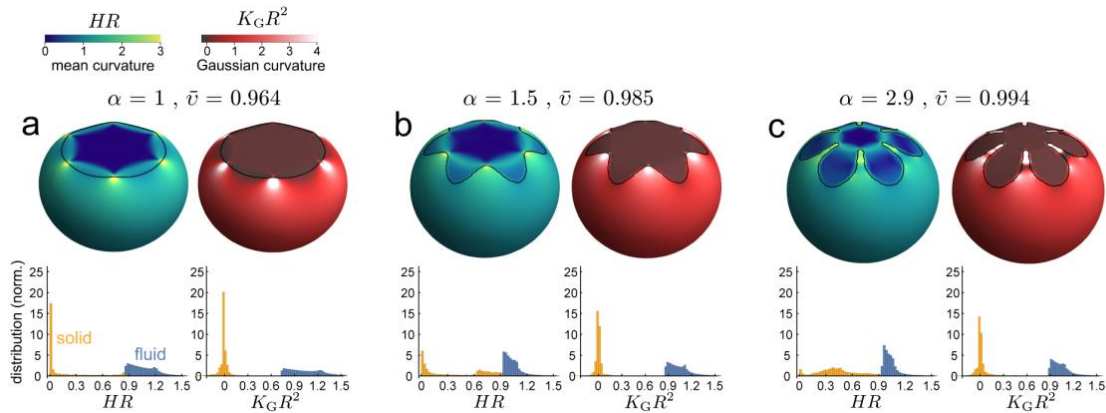

**Figure 7**. Images of three equilibrated vesicles, calculated with Surface Evolver, bearing a) circular ($\alpha = 1$), b) starfish ($\alpha=1.5$), and c) flower ($\alpha=2.9$) shaped solid domains, occupying 13% membrane area. Each vesicle is represented twice to show separate color scales for mean (blue green) and Gaussian (red black) curvature. Below each, a histogram quantifies the mean and Gaussian curvature in the fluid and solid membrane phases, normalized to 1. The vesicles are shown with the reduced volume, $\bar{v}$, the actual volume normalized on that of a sphere of the same area, selected to produce the zero tensioned-state that occurs just before further inflation produces membrane tension. This reveals that with the flower-shaped crystal, a greater volume of water can be accommodated before the tension rises. Conversely with the circle shaped crystal the volume that can be accommodated before the tension rises is somewhat smaller.

Figure 7. While line tension between the fluid and solid membrane phases always favors compact domain formation, the balance between line tension and bending costs, with the latter especially sensitive to large vesicle inflations, selects domain shape.[1] Though our prior publication revealed this mechanism in the context of bending energy, the membrane tension is a complimentary variable, the Legendre transform of the elastic energy at fixed volume. Thus the dependence of the preferred crystal shape on bending energy translates to a tension dependence.

Figure 7 provides insight into the preference for established crystals of different shapes; however, the current work demonstrates that the tension-driven preference for flower-shaped crystals is established by the tension build up in the one phase region even before nucleation of the solid phase. This finding suggests that membrane tension and bending exert a dominating effect even when the solid crystals are tiny, the scale of single microns on vesicles whose diameters are more than 10 times larger. At the same time, we cannot comment on, and do not necessary believe, that nucleation is itself fundamentally different when tension is increased though we have shown the nucleation density decreases with tension.[31] Indeed, we believe that all crystals in this study share the same molecular packing, based on their uniform exclusion of a series of tracer dyes,[1] some of which intercalate into $L_{\beta'}$ (gel) membrane phases in other systems.[9, 14, 50] The question of the impact of curvature on extremely small growing crystals and the possible morphological transition of submicron crystals to different shapes, by the time they are visible, remains an ongoing area of study.

A final point, the engineering model proves powerful in its ability to anticipate tensions that correlate with crystal shapes on vesicles of different sizes in Figure 5. However, the exact amount of cooling, from the point where a vesicle becomes spherical to when it enters the two phase region, has been estimated based on our observations of vesicle shape at elevated temperatures. There is variability in the relative fullness of different vesicles in a population, depending on how they pinched from an undulating lamella in the electroformer to produce vesicles, a process over which we have no control. The ranges of tensions shown, therefore are simply estimates within a working range. We were, however, quite pleasantly surprised to see how, well the tension values matched reasonable expectations for tensions based on lysis conditions and our measurements in Figure 6a.

**Summary and Outlook**

Single phospholipid crystals having 6-fold symmetry, nucleated and grown upon cooling within the vesicle membranes on the DPPC-dilute side of the phase diagram to avoid nucleation of multiple crystals, exhibit an increased tendency for protrusions and concave regions (flower shapes) with increased vesicle size. The strong vesicle size dependence, running counter to the size dependence of colloidal crystal growth on spherical templates, suggests the importance of membrane itself in controlling crystal morphology for crystals grown in vesicle membranes. In addition to the impact of vesicle size on crystal shape, we discovered a strong dependence on cooling rate, favoring more elaborate flower shapes for faster cooling. More critically, it was found that the cooling history preconditions the fluid membrane, even before nucleation occurs, to produce crystals of different shapes by the time they are visible. Once visible, the shapes of the tiny crystals are preserved during progressive growth during further cooling in the two phase region. Thus, for the range of conditions in this study which focus on single crystals in the DPPC-dilute part of the phase diagram, the fate of the crystal shape is set in the one phase region, even before nucleation.

The remarkable ability to precondition the stress of uniform fluid membranes to determine the morphology of subsequently nucleated crystals (by the time they are visible) is suggested by an engineering model for tension evolution, addressing the competing effects of thermal membrane contractions and water diffusion across the membrane. The membrane tension in large vesicles can become substantial even before the phase boundary is crossed, not the case for smaller vesicles. The model predicts a tension at the entrance to the two phase region, for vesicles of different size and cooling rate on the order of several mN/m and approaching lysis, correlating with the morphological boundaries in experimental state space. The impact of high tension to produce more elaborate flower shapes can be understood from the ability of flower petals to bend cylindrically to relax membrane stress in the solid phase; however, it was not anticipated that control of tension in the one phase region could predetermine crystal shape even before nucleation.

While stresses typically develop when materials are cooled in a variety of processing scenarios, and while surface area to volume ratio enables an intuitive understanding of why larger vesicles experience greater stresses to alter crystal shape, modeling revealed some surprising features. Most importantly is a qualitative difference in the tension build up in small versus large vesicles: For the range of properties in fluid membranes and for cooling rates that are practical in a lab time frame, small vesicles, below about 25 μm for a cooling rate of 0.30 °C/min, experience a low steady state (constant) tension that enables the growth of compact domains without a large bending energy cost. The tension in larger vesicles is, however, not limited by an early steady state plateau and instead grows substantially with increased cooling, favoring crystals to take on flower shapes by the time they are large enough to be imaged in a fluorescence microscope. The cooling rate can be adjusted to tune the interplay between size and tension, with very slow cooling rates enabling mostly compact crystals over a large range of vesicle sizes.

The findings in this study emphasize the utility of thinking about biomolecular membranes in the same context as we would classical materials, for instance polymers, resins, fibers, and films where processing rates influence stress development, ultimately enabling control of morphological and performance related properties. The ability to precondition vesicles in suspension through controlled processing rates may facilitate the use of biomolecular and biomimetic membranes as platforms to scalably produce valuable 2D materials for energy applications. Further the new perspective on membrane tension evolution in vesicles of different sizes, provided by the engineering model, may enable rational design of vaccines and other biomolecular technology.

**Materials and Methods**

1,2-dioleoyl-sn-glycero-3-phosphocholine (DOPC), 1,2-dipalmitoyl-sn-glycero-3-phosphocholine (DPPC), and tracer lipid 1,2-dioleoyl-*sn*-glycero-3-phosphoethanolamine-N-(lissamine rhodamine B sulfonyl) (ammonium salt) (Rh-DOPE) were purchased from Avanti Polar Lipids (Alabaster, AL). A solution containing 31 mol% DPPC, 69 mol% DOPC, and < .1 mol% fluorescent tracer lipid, with a total lipid concentration of 5 mg/ml in chloroform (Sigma Aldrich) was placed dropwise on the platinum wires of the electroformer. After drying, the electroform was sealed and filled with a 10 mM sucrose solution that had been preheated to 60°C. The chamber was heated in the range 55-70 °C in the one phase region to ensure uniform composition of all vesicles in a batch, and a current 3V and 10 Hz was applied for an hour, after which time, the vesicle suspension was harvested in a syringe and stored at room temperature.

To study crystal morphology at controlled conditions, vesicles were diluted 10- fold in DI water and annealed near 55°C in a custom built chamber, whose temperature was controlled by flowing water from a temperature control bath, to melt any solids and then quickly brought to a temperature in the range 42-45°C. From this point, a cooling program having the rates specified in the main paper, was implemented. Cooling was best controlled when the sealed chamber was insulated, enabling viewing of vesicles only after reaching room temperature. However, in select experiments, the chamber was placed on the microscope stage to observe the evolving crystal shapes. Imaging was conducted employing a 40x objective on a Nikon Eclipse TE300 epifluorescence microscope. Images were recorded with a pco.panda 4.2 sCMOS monochrome camera with a resolution of 0.17 μm / pixel.

From a thermodynamic model, program Surface Evolver was employed to compute the shapes and bending energies of equilibrated vesicles of volume $V$ containing solid membrane domains having an overall membrane area fraction, $\phi = 0.13$, with the remaining fluid membrane area $1 - \phi$. The hexagonally symmetric cutout shapes of the solid domains were chosen to match the experimental observations of round, starfish, or flowers. The solid domains were modeled as elastic plates with 2D Youngs modulus, $Y$, Poisson ratio $v$, and plate bending modulus $B$:

$$\frac{Y}{2(1+v)} \int_{\text{solid}} dA \left[ (\text{Tr } \varepsilon)^2 + \frac{v}{1-v} \text{Tr } \varepsilon^2 \right] + \frac{B}{2} \int_{\text{solid}} dA (2H)^2 \qquad (8)$$

where $\varepsilon$ is the 2D strain and $H$ is the mean curvature. The fluid membrane, with its fixed area, was treated using the Helfrich model, with a bending energy $E_{\text{fluid}} = \frac{B}{2}\int_{\text{fluid}}(2H)^2$. Surface Evolver calculations requires setting finite values of for $Y$ and $B$ (we use $\nu = 0.4$), which we set by the ratio defined by their ratio $t = \sqrt{B/Y}$, which defines the *elastic thickness* of the solid. We expect this ratio to be of order of the few nanometer thickness of the bilayer, which is made dimensionless by comparison to the radius multi-micro of the composite vesicle $R$. For the simulations reported here we set $\frac{t}{R} = 1.5 * 10^{-4}$ which implies that in-plane strains elastically prohibitive relative to bending deformations.

Employing the same value of B for the fluid and solid for simplicity produces the lower-bound estimate for the impact of solid stiffness, demonstrating it is not the solid's resistance to (cylindrical) bending that is responsible for the energy cost but rather the solid shear rigidity captured in the first term of the integral, that distinguishes the solid. Because Gaussian curvature is forced into the membrane fluid and because the vesicle tangent remains continuous at the fluid-solid boundary bending terms from the Helfrich energy that couple to the Gaussian curvature in both fluid and solid phases are independent of the vesicle shape. The Gaussian curvature, $K_G$, integrated over the entire vesicle must remain constant at $4\pi$, and with $K_G$ vanishing in the solid, $\int_{\text{fluid}} dA\, K_G = 4\pi$ for all configurations. While $K_G$ is not strictly constrained to vanish in the solid, we have confirmed that the in-plane solid elasticity leads to near-perfect expulsion of Gaussian curvature from the solid,[51] also shown in Figure 7. Surface evolver was employed to consider variations in vesicle shape (both the bending of the fluid and solid membrane regions) to minimize the integrated energy at fixed vesicle volume and fixed fluid and solid area, producing the shapes of Figure 7. Extensive details can be found in prior work.[1]

## Acknowledgements

This work was supported by DOE DE-SC0017870

Supporting Information:

**Thermal Preconditioning of Membrane Stress to Control the Shapes of Ultrathin Crystals**

Hao Wan,[1] Geunwoong Jeon,[2] Gregory M. Grason,[1] Maria M. Santore[1,*]

[1]Department of Polymer Science and Engineering
University of Massachusetts
120 Governors Drive
Amherst, MA 01003

[2]Department of Physics
University of Massachusetts
710 N Pleasant St
Amherst, MA 01003

*corresponding author: santore@mail.pse.umass.edu

1. **Distribution of vesicle sizes for different crystal shapes, measured at different cooling rates.**

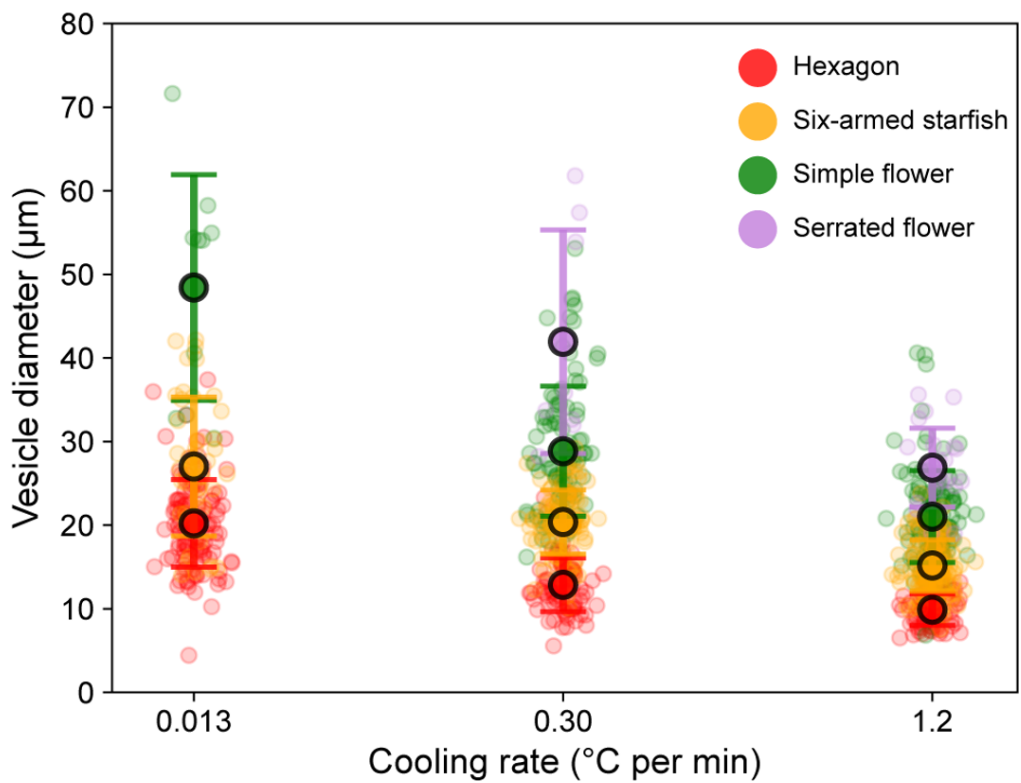

## 2. Impact of membrane property parameters on tension evolution, from the model.

Below we show the impact of membranae properties on the tension evolution for an hour of cooling at the rate of 0.3 °C/min. The findings generally show that despite substantial variations in properties, the small vesicles, 20 μm in diameter or less, reach a steady state plateau that varies only slightly with the membrane property of interest. Therefore the impact of the membrane property on the plateau level and timing of the plateau would be very difficult to distinguish in experiments with small vesicles.. The greatest impact of membrane properties is seen for the largest vesicles, approaching 60 μm in diameter. Here the rapid rise in tension was sensitive to membrane properties but, because the large tensions approached lysis conditions, burst and reseal dynamics may dominate in for the largest vesicles.

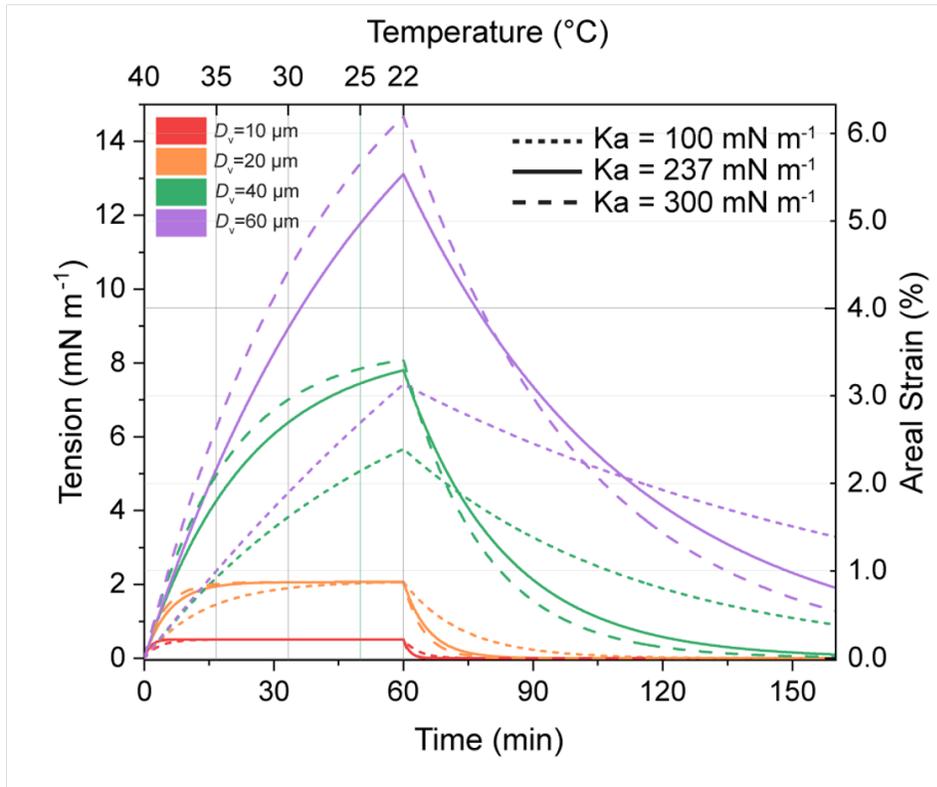

**Figure S-2.** Impact of $K_a$, area expansion modulus, on tension evolution during cooling and subsequent temperature hold near room temperature. Each color family represents a vesicle size while the line types show variations in $K_a$. Similar to the main paper, the upper axis estimates the temperature history for the case where vesicles become spherical and first develop tension at 40°C.

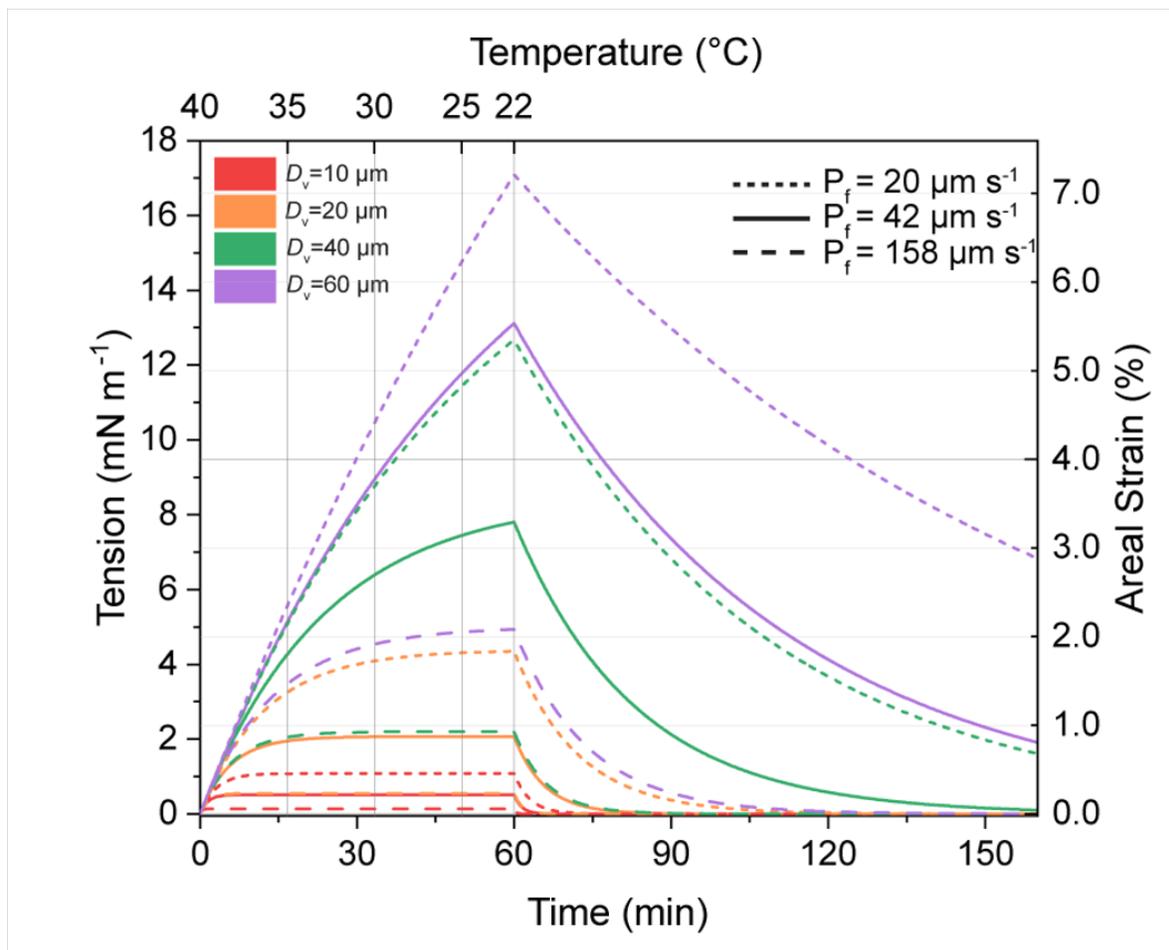

**Figure S-3.** Impact of membrane permeability, on tension evolution during cooling and subsequent temperature hold near room temperature. Each color family represents a vesicle size while the line types show variations in permeability. Similar to the main paper, the upper axis estimates the temperature history for the case where vesicles become spherical and first develop tension at 40°C.

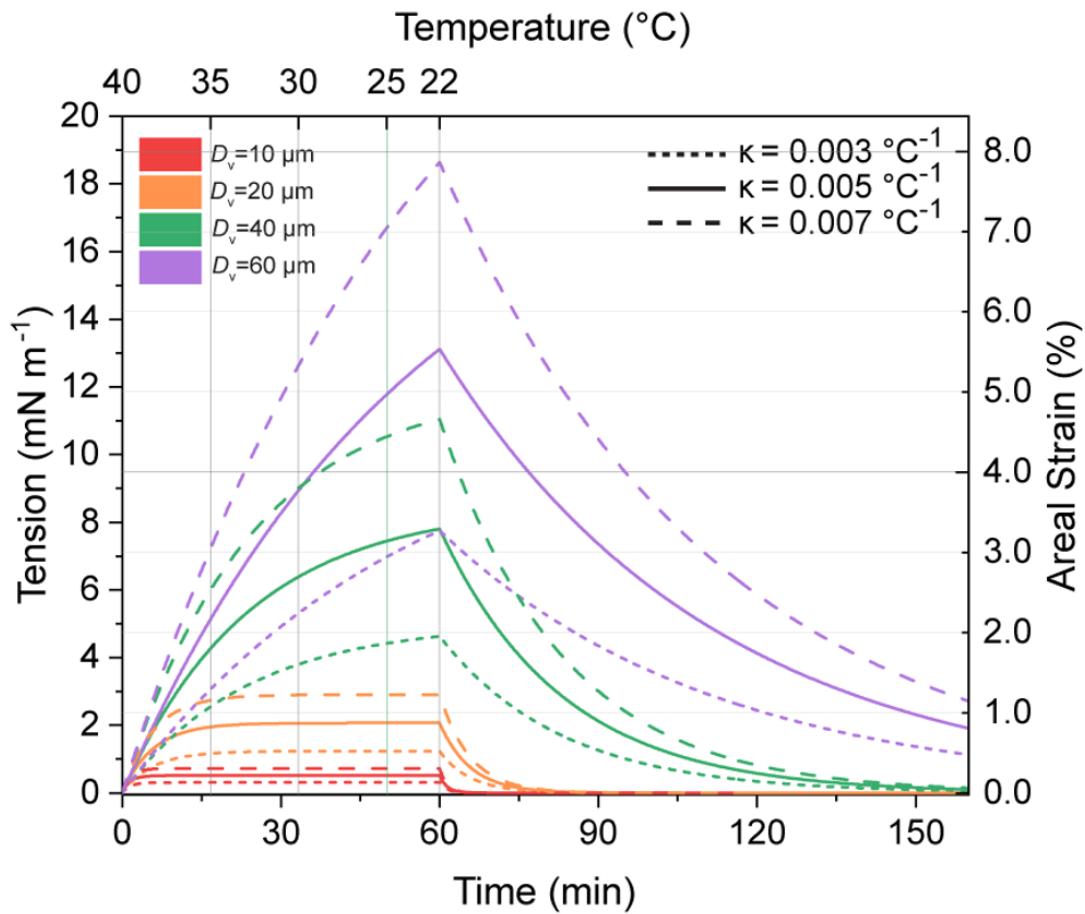

**Figure S-4.** Impact of coefficient of thermal expansion, on tension evolution during cooling and subsequent temperature hold near room temperature. Each color family represents a vesicle size while the line types show variations in coefficient of thermal expansion. Similar to the main paper, the upper axis estimates the temperature history for the case where vesicles become spherical and first develop tension at 40°C.